\documentclass[fleqn,usenatbib]{mnras}

\usepackage{newtxtext,newtxmath}

\usepackage[T1]{fontenc}
\usepackage{natbib}
\DeclareRobustCommand{\VAN}[3]{#2}
\let\VANthebibliography\thebibliography
\def\thebibliography{\DeclareRobustCommand{\VAN}[3]{##3}\VANthebibliography}

\usepackage{graphicx}	
\usepackage{newtxtext,newtxmath}
\usepackage[T1]{fontenc}
\usepackage{ae,aecompl}
\usepackage{epsfig}
\usepackage{amsmath, amsfonts, epsfig, xspace}
\usepackage{natbib}
\usepackage{deluxetable}
\usepackage{rotating}
\usepackage{graphicx}
\usepackage{caption}
\usepackage{longtable}
\usepackage{multicol}
\usepackage{booktabs}
\usepackage{dcolumn}
\usepackage{longtable}
\usepackage{changepage}
\usepackage{comment}
\usepackage{pdflscape}

\title[INOV of ZTF Blazars and RQQs]{Intranight Optical Variability of Blazars and Radio-quiet Quasars using the ZTF Survey}
\author[Vibhore Negi et al.]{
Vibhore Negi,$^{1,2}$\thanks{E-mail: vibhore.negi18@gmail.com}
Gopal-Krishna,$^{3}$
Ravi Joshi,$^{4,5}$
Hum Chand,$^{6}$
Paul J.\ Wiita,$^{7}$
Navaneeth P K,$^{8}$
\newauthor
~~Ravi S. Singh$^{2}$
\\
$^{1}$Aryabhatta Research Institute of Observational Sciences (ARIES), Manora Peak, Nainital, 263002, India\\
$^{2}$Department of Physics, Deen Dayal Upadhyaya Gorakhpur University, Gorakhpur, 273009, India\\
$^{3}$UM-DAE Centre for Excellence in Basic Sciences, Vidyanagari, Mumbai, 400098, India\\
$^{4}$Indian Institute of Astrophysics, Koramangla, Bengaluru, 560034, India\\
$^{5}$Kavli Institute for Astronomy and Astrophysics, Peking University, Beijing 100871, China\\
$^{6}$Department of Physics and Astronomical Science, Central University of Himachal Pradesh (CUHP), Dharamshala, 176215, India\\
$^{7}$Department of Physics, The College of New Jersey, PO Box 7718, Ewing, NJ, 08628-0718, USA\\
$^{8}$Central University of Karnataka, Kadaganchi, 585367, India \\
}
\date{Accepted XXX. Received YYY; in original form ZZZ}

\pubyear{2022}

\begin{document}
\label{firstpage}
\pagerange{\pageref{firstpage}--\pageref{lastpage}}
\maketitle
\begin{abstract}
We explore the potential of the ongoing Zwicky-Transient-Facility (ZTF) survey for studying Intra-Night Optical Variability (INOV) of active galactic nuclei (AGN), in particular for picking rare events of large INOV amplitudes, whose detection may require extensive temporal coverage. For this, we have used the available high cadence subsets of the ZTF database to build a well-defined large sample of 53 blazars (BLs) and another sample of 132 radio-quiet quasars (RQQs), matched to the blazar sample in the redshift$-$magnitude plane. High-cadence ZTF monitoring of these two matched samples are available, respectively, for 156 and 418 intranight sessions. Median durations for both sets of sessions are 3.7 hours. The two classes of powerful AGN monitored in these sessions represent opposite extremes of jet activity. The present analysis of their ZTF light curves has revealed some strong INOV events which, although not exceptionally rare for blazars, are indeed so for RQQs, and their possible nature is briefly discussed.
\end{abstract}




\section{Introduction}

\begin{figure*}
	\includegraphics[width=1.0\textwidth,height=0.40\textheight]{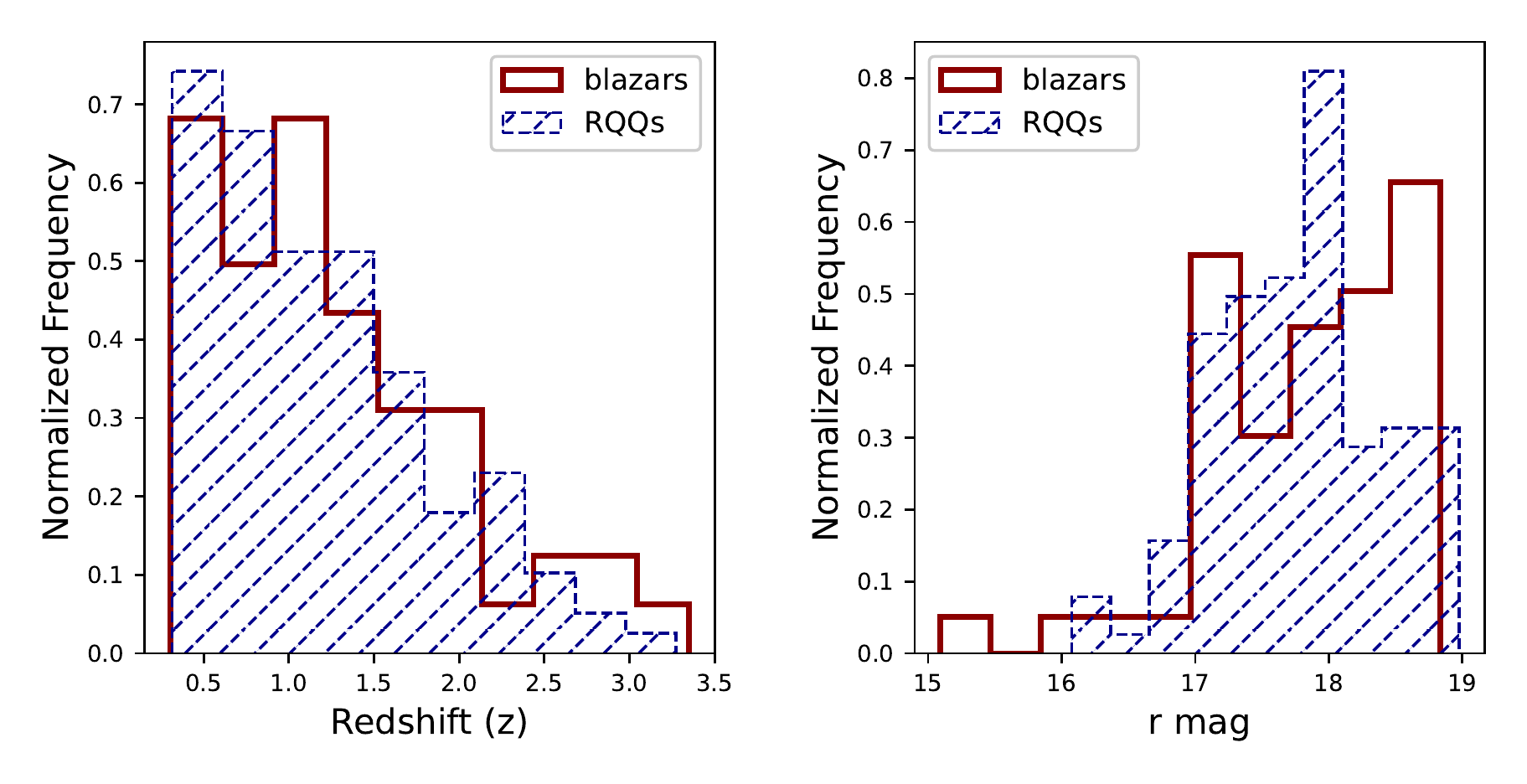}
    \caption{The redshift and apparent magnitude distributions of the sample of blazars (53) and the control sample of RQQs (132). {\textit {Left panel}}. Median values of redshift for the two distributions are 1.15 (blazars) and 1.07 (RQQs).
    {\textit {Right panel}}: Distribution of the r-band apparent magnitudes for the blazar and RQQ samples, having median values of 18.0-mag and 17.8-mag, respectively.}
    \label{fig:sample_distribution}
\end{figure*}

Active Galactic Nuclei (AGNs) are known to radiate over a very wide
frequency range, from radio to TeV $\gamma$-rays, and to show flux variability across the electromagnetic spectrum on diverse time scales \citep{1978bllo.conf..328B,1995PASP..107..803U, 1995ARA&A..33..163W, 1997ARA&A..35..445U}. This characteristic has been extensively used to probe the physics of their central engines which remain spatially unresolved in direct observations \citep[e.g., reviews by][]{1993ARA&A..31..473A, 2016Galax...4...37M, 2019ARA&A..57..467B}. The optical flux variations of AGNs on hour-like (or, even shorter) time scales are commonly referred to as ‘micro-variability’ \citep{1989Natur.337..627M} or ‘Intra-Night Optical Variability’ (INOV, \citealt{1995MNRAS.274..701G,2003ApJ...586L..25G}). In the case of blazars and other $\gamma$-ray detected AGN, INOV amplitude ($\psi$) reaching a tenth of magnitude, or even more, is not very rare, whereas much weaker INOV is displayed by the remaining classes of AGN \citep[e.g.,][hereafter GW18]{2004JApA...25....1S, 2004MNRAS.350..175S, 2013MNRAS.435.1300G, 2018BSRSL..87..281G}. \par
The INOV of blazars is widely linked to turbulence in their relativistic jets, \citep[e.g.,][]{1992vob..conf...85M, 2012A&A...544A..37G, 2014ApJ...780...87M, 2015JApA...36..255C, 2021Galax...9..114W}, such that any tiny emissivity fluctuations arising within the jet flow can appear hugely amplified due to the jet’s bulk relativistic motion \citep[e.g.,][]{2009MNRAS.395L..29G, 2012A&A...548A.123B, 2012MNRAS.420..604N, 1985MNRAS.215..383S}. It has been suggested that a broadly similar but milder version of this process, occurring inside intrinsically weak and less well aligned relativistic jet, could account for the low-level INOV exhibited by RQQs (\citealt{2003ApJ...586L..25G}, hereafter GSSW03; \citealt{2004MNRAS.350..175S}). On the other hand, INOV of RQQs could even arise from flares occurring in the magnetised plasma of the accretion disk and in its environment \citep{1993ApJ...406..420M, 1991A&A...246...21Z}, or due to shocks/instabilities in the accretion disk \citep{1993ApJ...411..602C}. Such possibilities make INOV studies a useful probe of the physics of the central engine and the relativistic jets emanating from it. 
\par

Although a good deal of observational effort has been invested over the past few decades, on characterising the INOV displayed by diverse classes of AGN, the focus in most campaigns has clearly been on blazars (e.g., see the reviews GW18; \citealt{2018Galax...6....1G, 2021Galax...9...69W}).
In such studies, each AGN was typically monitored quasi-continuously for durations between 2 and 6 hours, using 1 - 2 metre class optical telescopes, thus achieving an INOV detection limit of $\sim$ 1 - 2\% of amplitude ($\psi$). A large duty cycle of INOV detection (DC $\sim$ 50-70\%) has thus been established for BL Lacs and the flat-spectrum radio quasars (FSRQs) exhibiting a high fractional polarisation ($p_{opt}$ > 3\%). In contrast, the remaining AGN classes, even including low$-$polarisation FSRQs (LPQs) whose nuclear radio jets are relativistically beamed towards us, and the lobe-dominated radio quasars (LDQs), were all found to exhibit only mild INOV (typically, $\psi$ $\lesssim$ 3\%), with a small INOV DC of around 10\% \citep[see, e.g., GSSW03 and references therein;][GW18]{2013MNRAS.435.1300G, 2004JApA...25....1S, 2002A&A...390..431R}. \par
In this work we report a search for high-amplitude INOV of blazars and RQQs, the two classes of powerful AGN that represent the opposite extremes in jet activity. As compared to their previous INOV studies, our present samples of these two AGN classes are significantly larger and have been monitored more extensively, as part of the ZTF survey \citep{2019PASP..131a8003M}, without any consideration of the activity levels of the individual sources. The ZTF survey scans the northern sky using a 47 deg$^{2}$ wide-field imager mounted on a 48-inch Schmidt telescope on Mount Palomar. The database provides long-term light curves interspersed with high cadence light-curves (LCs) which can be used for detecting rare events of large INOV. We first describe in Section~\ref{sec:data_sample} our sample selection and the ancillary data used. Section~\ref{sec:analysis_results} describes the data analysis and presents the results obtained. This is followed by a brief discussion in Section~\ref{sec:discussion}. Our main conclusions are summarized in Section~\ref{sec:conclusion}.

\section{The Data and Sample Selection}
\label{sec:data_sample}
\begin{figure}
	\includegraphics[width=0.5\textwidth,height=0.35\textheight]{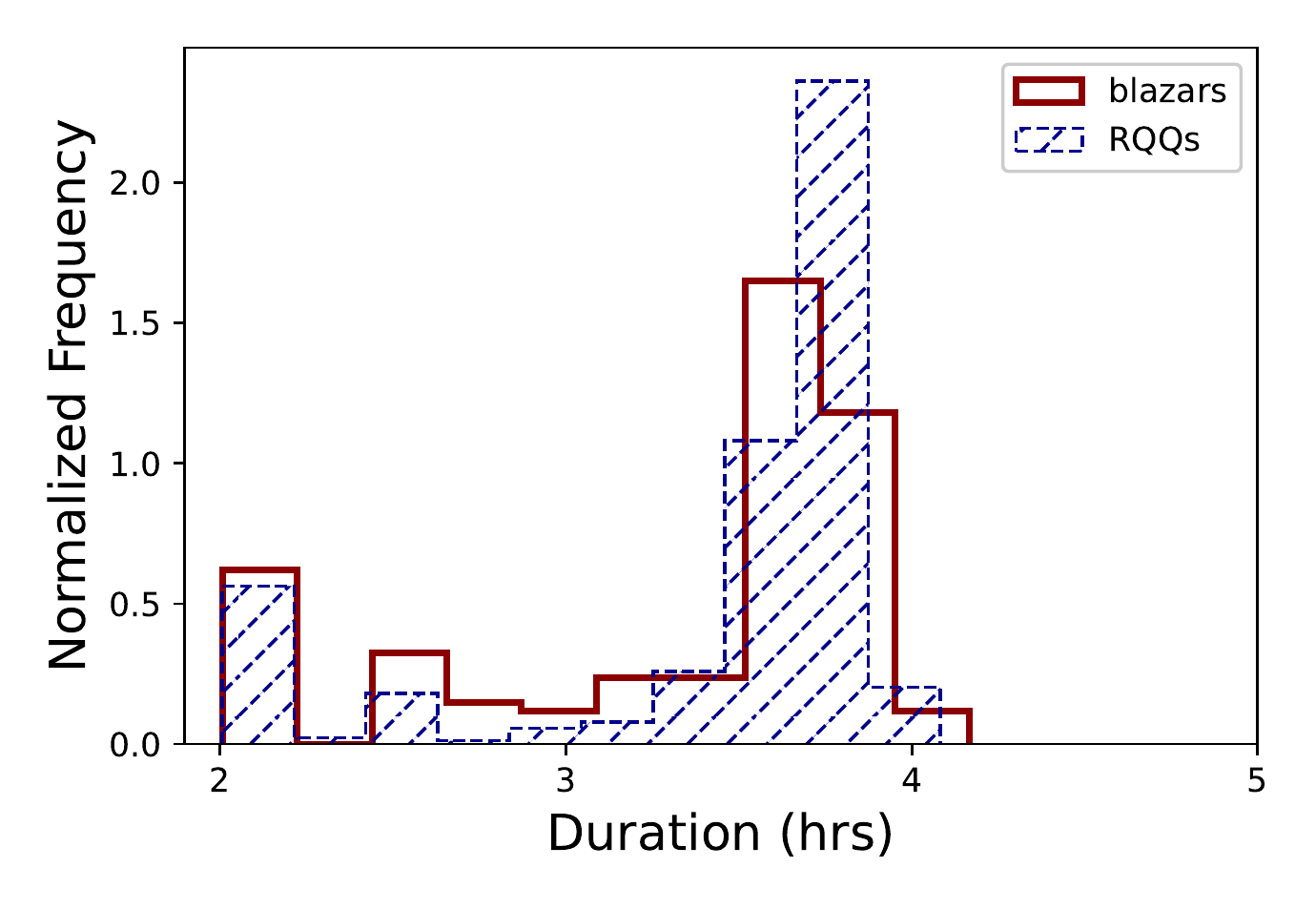}
    \caption{The distributions of the observed time duration (T) for the light-curves of blazars (156) and RQQs (418). The median values for both histograms is 3.7 hours.}
    \label{fig:sample_dur_distribution}
\end{figure} 
The parent catalogue of the blazars covered in this study is the ROMA-BZCAT \citep{2015Ap&SS.357...75M}, which lists a total of 3561 blazars, classified as BL Lacertae objects (BL Lacs), or flat-spectrum radio quasars (FSRQs). Only 2751 of them were found in the ZTF database (ZTF DR10\footnote{\url{https://www.ztf.caltech.edu/ztf-public-releases.html}}). Out of these, we discarded 159 blazars labelled as `uncertain type’ and another 71 as ‘candidate’ BL Lac in the ROMA-BZCAT. Next, since the surface-brightness profile of the host galaxy of an AGN does not generally behave like a point source when convolved with the `point-spread function’ (PSF), any intranight variation in PSF (i.e., `seeing') can give rise to a changing ratio of light contributions from the blazar and its host galaxy. Such light-curves (LCs) can lead to spurious claims of  INOV \citep{1991AJ....101.1196C,2000AJ....119.1534C}. In order to reduce this possibility to a negligible level, we have only selected sources lying at redshifts $z$ > 0.3, leaving us with 1555 sources. Their (PSF-fitting based) r-band LCs available in the ZTF database were firstly cleaned up by accepting only the `good quality’ data, i.e., discarding the data points with ZTF CATFLAG score of zero. The ZTF survey determines model PSFs and employs them to generate a PSF-fit photometry catalog using a version of DAOPhot (\citealt{1987PASP...99..191S}) optimized for ZTF, including an added benefit of de-blending by fitting to multiple detections simultaneously in a two-pass PSF fit \citep{2019PASP..131a8003M}.
Further, ZTF treats the LCs observed in a particular field, filter, and CCD-quadrant independently, and assigns them different observation IDs. Since, combining the LCs from different fields and CCD-quadrants can induce spurious variability, we have accepted for a given source only the LC corresponding to the observation ID with the highest number of data points \citep[see also,][]{2022MNRAS.510.1791N}. Next, in order to reduce the number of LCs with poor signal to noise ratio, we rejected all the sources having mean $m{_r}$ > 19 over the duration of the ZTF survey. \par

As further filters, we rejected the LCs of duration (T) < 2 hours and also those having gap(s) of > 1 hour.  As ZTF employs a small telescope (48-inch) and the exposure time for a single frame is small ($\sim$30 sec), the photometric errors are relatively large for fainter sources. Hence we have carried out a 3-point median binning of the LCs, when required, provided the binning does not reduce the number (N) of data points in a LC to below 10. Thereafter, in order to reduce noisy LCs, we rejected sessions for which average quoted photometric error of the data points is  > 0.03 mag. \par
Lastly, as our final filter, all remaining sessions with less than 10 data points were discarded, leaving us with a final sample of 53 blazars (BLs), monitored in 156 ZTF intranight sessions. These consist of 46 FSRQs (138 sessions) and 7 BL Lacs (18 sessions). For comparing the INOV of this sample of BLs with that of RQQs, we then constructed a preliminary `matched sample’ of RQQs out of the SDSS-DR14 \citep{2018A&A...613A..51P}. The matching was done in the $z - m{_r}$ plane, setting tolerances of $\pm 0.1$ and $\pm 0.2$-mag, respectively. This resulted in a set of about 20 matched RQQs for each BL. For the matched sample of RQQs, we checked for availability of ZTF r-band LCs and the same were found for 132 of the RQQs (418 sessions). KS-test on the two matched samples provides a D-statistic of 0.18 for magnitude and 0.09 for redshift, with a probability of null hypothesis being  0.14 and 0.91, respectively (see Fig~\ref{fig:sample_distribution}). Details of the BL sample and the matched RQQ sample are given in Tables~\ref{tab:sample_table_BL} and \ref{tab:sample_table_RQQ}. The median durations (T) of the LCs are found to be 3.7 hours for both the BL sample and the matched RQQ sample, with $\sim$75 per cent of the BL sessions and $\sim$84 per cent of the RQQ sessions having T > 3 hours (up to $\sim 4$ hours, see Fig.~\ref{fig:sample_dur_distribution}).
\section{Analysis and Results}
\label{sec:analysis_results}
A preliminary check for INOV in the ZTF LCs (which are based on PSF-fitting photometry) of our matched samples of 53 blazars (156 sessions) and 153 RQQs (418 sessions), was made employing the widely used {\it F}-test \citep{2010AJ....139.1269D}. The {\it F}-value for each LC was computed from: \begin{equation} 
\label{eq:ftest2}
F = \frac{Var(q)}
{ \sum_\mathbf{i=1}^{N}\sigma^2_{i,err}(q)/N}
\end{equation}
where $Var(q)$ is the variance of the LC of the target AGN (RQQ/BL), $\sigma_{i, err}(q)$ is the rms error on the $i^{th}$ data point in the LC of the target AGN, and $N$ is the number of data points in the LC. The computed value of $F$ for each LC was compared with the critical $F$-value ($F_{c}$) computed for that session, for 99\% confidence level. The LCs showing $F \geq$ $F_{c}$ (0.99) were shortlisted as candidates for INOV detection. In the next level check, each candidate LC was subjected to visual inspection of the optical field around both the target AGN and 3 - 4 comparison stars monitored concurrently with the AGN on the same CCD chip. The comparison stars were selected applying the criteria of a clean optical neighbourhood (the criterion also employed for the AGN, before claiming its INOV detection), as well as proximity to the target AGN in both position (within $\sim$ 5 arcmin) and $m{_r}$ (|$\Delta m{_r}$ | $\lesssim$ 1.0). The LCs of the comparison stars were then subjected to the $F$-test mentioned above. Applying this procedure, which is equivalent to the widely practiced `differential photometry’, to the short-listed LCs of both the target AGN and the comparison stars, we were able to isolate good INOV candidate LCs of the blazars and RQQs (for which the computed $F > F_{c}(0.99)$, unlike the set of comparison stars). All these LCs were subjected to a final check for any systematic variation in PSF (obtained from the ZTF database) during the respective monitoring session and any AGN LC showing a variability trend correlated with the PSF variation was deemed to be unreliable and discarded. The above sequence of reliability checks has led to confirmation of INOV for 6 blazar sessions (involving 4 blazars) and 3 RQQ sessions. \par
\begin{landscape}
\begin{figure}

\begin{center}
\noindent\includegraphics[scale=0.5]{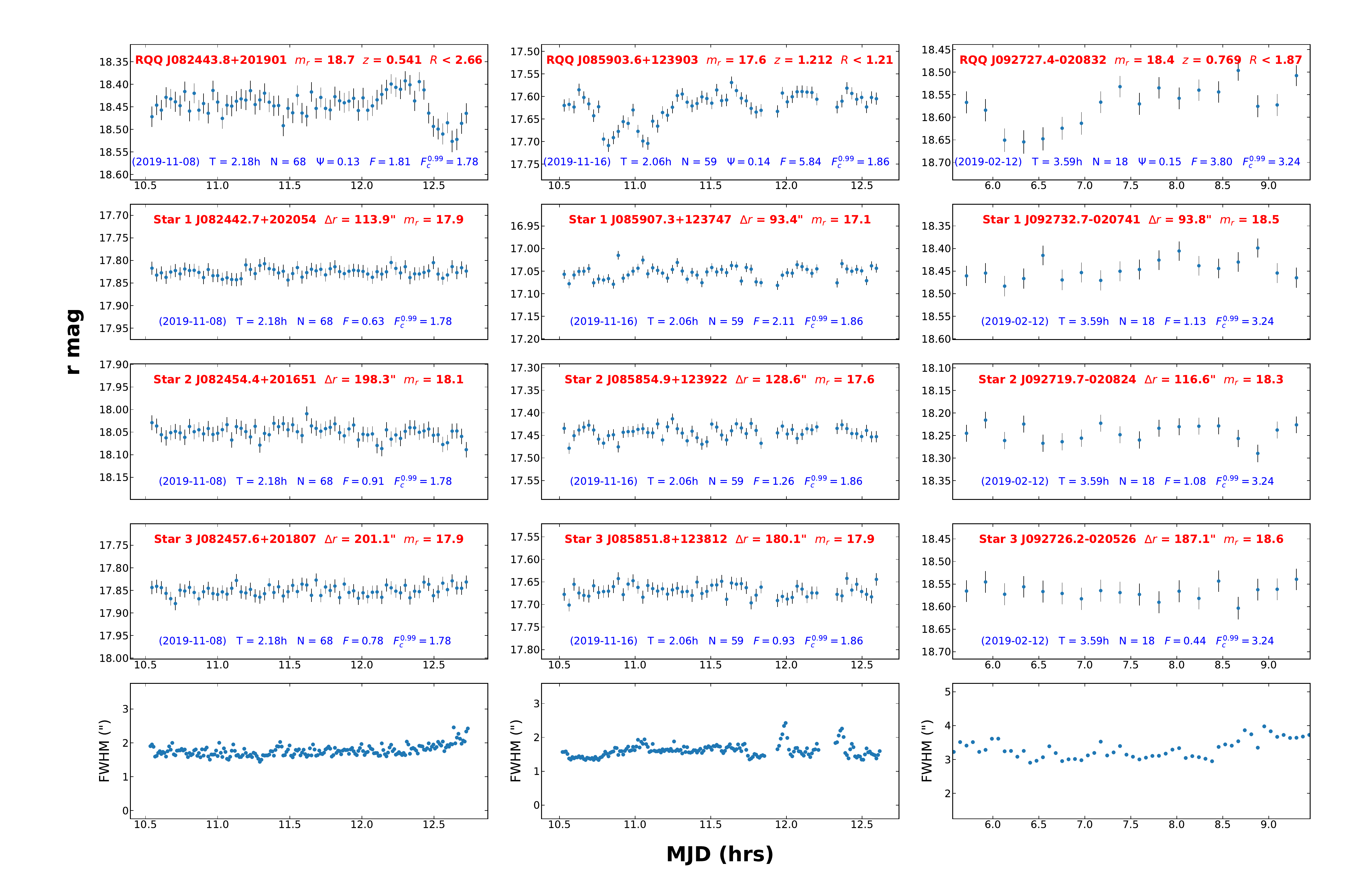}
    \caption{The LCs of the three RQQ sessions with INOV confirmed in the present analysis. The top panel shows the LC of the RQQ and the subsequent three lower panels show the LCs of 3 comparison stars in the order of increasing distance ($\Delta r$) from the RQQ. For each RQQ/star, the basic properties are given near the top of the panel, whereas the estimated parameters from the variability analysis for the session are displayed near the bottom of the panel. The lowest panel shows the variation of the seeing (PSF) through the monitoring session.}
    \label{fig:light_curve_RQQ_1}
\end{center}
\end{figure}
\end{landscape}

Table~\ref{tab1:var_result} lists the computed $F$ values for all these 9 LCs found to exhibit INOV in our analysis, along with the critical $F$-values at the desired significance level of ${\alpha}$ ($=F_{c}^{\alpha}$). 
All these 9 AGN LCs, together with the LCs of the corresponding comparison stars are displayed in Figures~\ref{fig:light_curve_BL_1}-\ref{fig:light_curve_RQQ_1}, together with the respective PSF variation profiles. The images of the optical fields are shown in Figures~\ref{fig:final_blazar_image_cutouts} and \ref{fig:final_rqq_image_cutouts_2}.
Also provided with each AGN LC are several observational and computed parameters, including the $F$-values (Eq.~\ref{eq:ftest2}) and the INOV amplitude ($\psi$) which has been computed from the relation \citep{1998A&A...329..853H}: 
\begin{equation} 
\label{eq:Psi_relation}
\psi= \sqrt{({A_{max}}-{A_{min}})^2-2\sigma^2}
\end{equation}
where $A_{max}$ and $A_{min}$ are the maximum and minimum values of the AGN LC and $\sigma^2=<\sigma^2_{q}>$, where, $\sigma_{q}$ is the mean rms error for the data points in the LC. \par
\begin{landscape}
\begin{figure}
\includegraphics[scale=0.5]{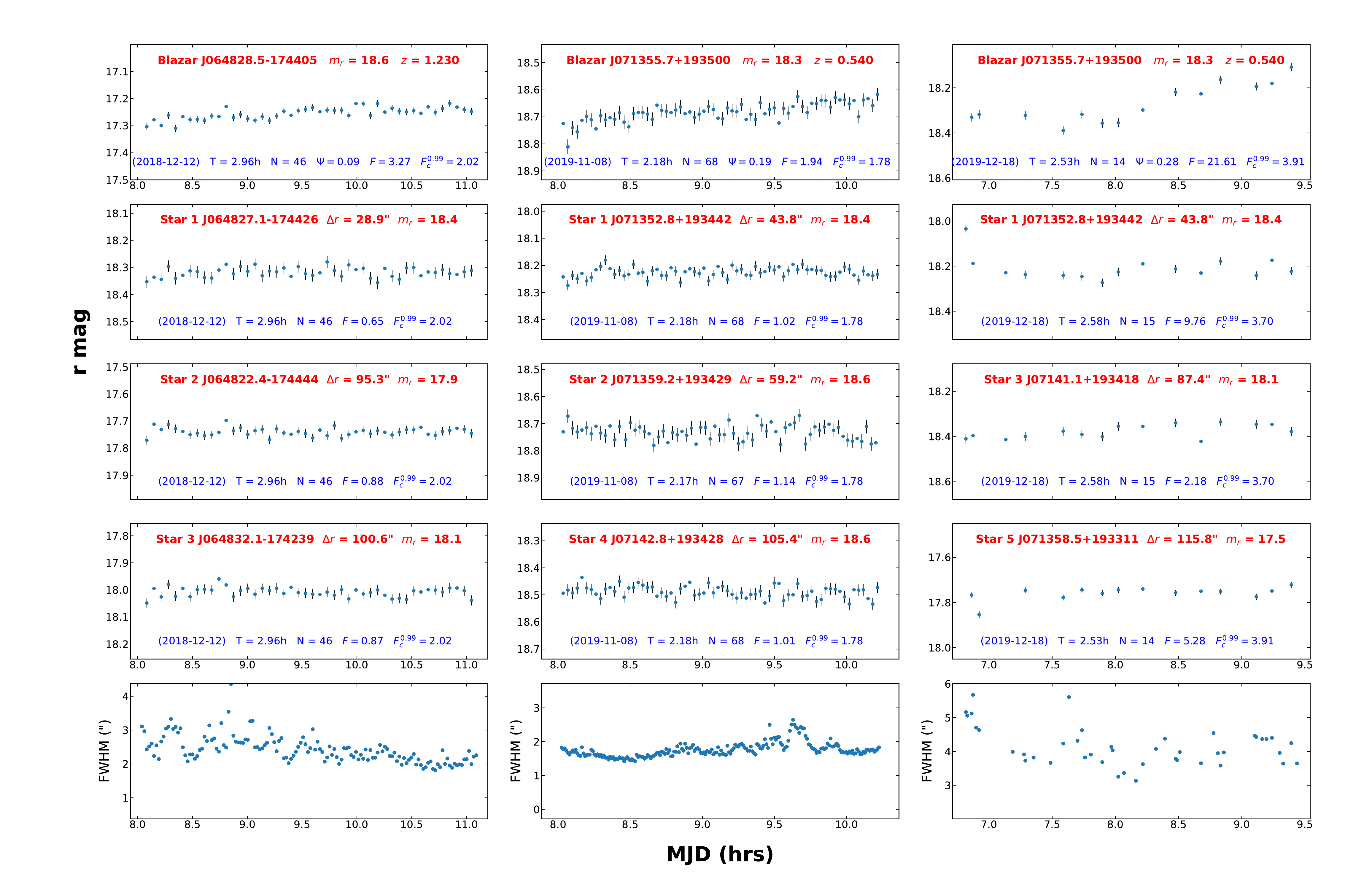}
    \caption{ The light-curves (LCs) of the blazar sessions with INOV confirmed in the present analysis. For each session, the top panel shows the LC of the target blazar, and the subsequent three lower panels show the LCs of the 3 selected comparison stars, in order of increasing distance ($\Delta r$) from the blazar. For each blazar/star, basic properties are mentioned near the top in the panel, whereas the estimated parameters from the variability analysis of the session are displayed near the bottom of the panel. The lowest panel shows the variation of the seeing (PSF) through the monitoring session.}
    \label{fig:light_curve_BL_1}
\end{figure} 
\end{landscape}
\begin{landscape}
\begin{figure}
\centering
\includegraphics[scale=0.5]{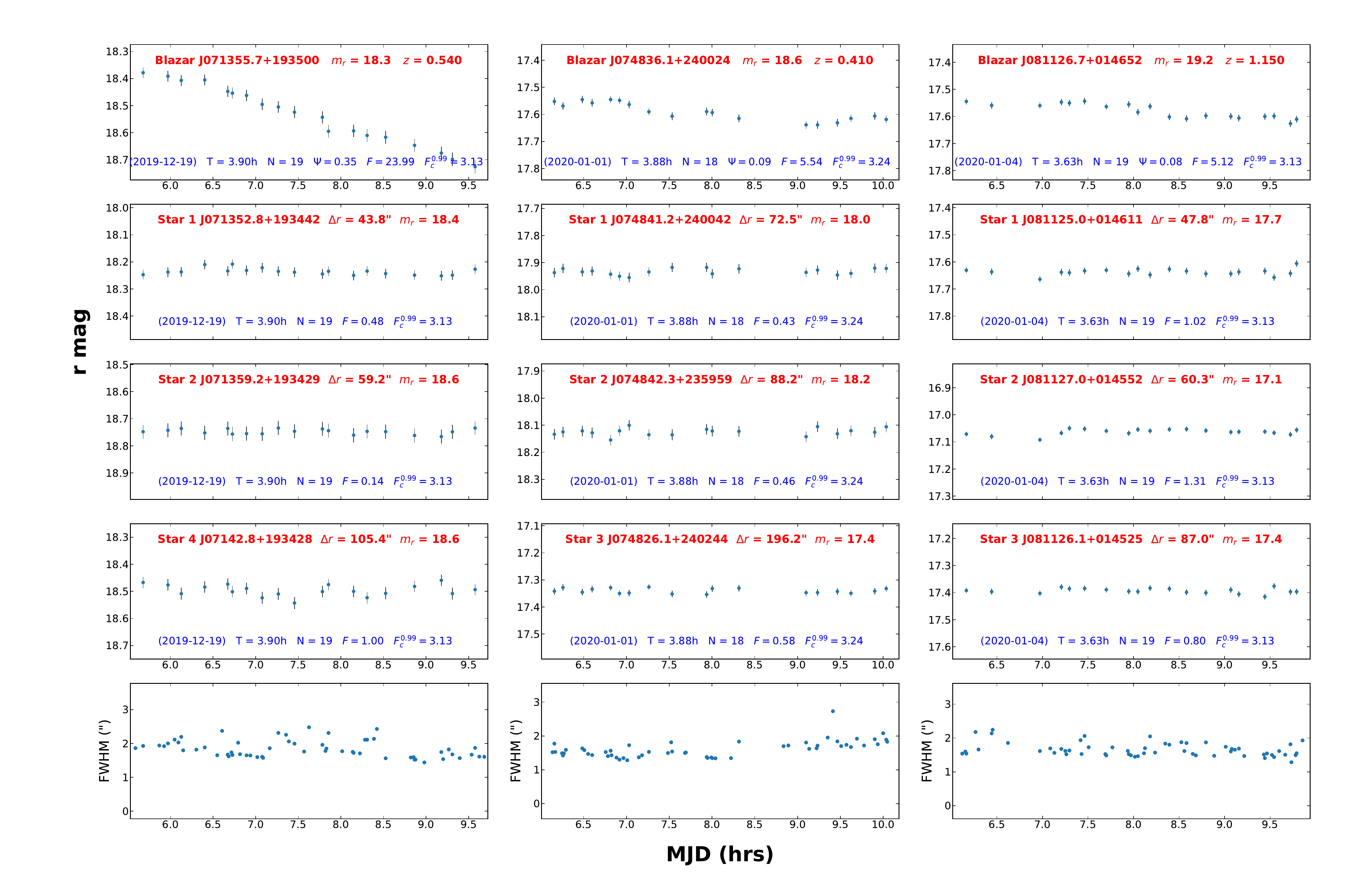}
    \setcounter{figure}{2}
    \caption{ continued.}
    \label{fig:light_curve_BL_2}
\end{figure} 
\end{landscape}

\begin{table*}
\begin{adjustwidth}{3cm}{}
\begin{minipage}{400mm}    
 \caption{Results of INOV analysis for the blazars and RQQs in our sample, showing INOV detection.}
    \hspace{-0.2in}
 \begin{tabular}{@{}cccccccc@{}} 
    \hline
 \multicolumn{1}{c}{AGN}
 &\multicolumn{1}{c}{Date of Obs.}
 &\multicolumn{1}{c}{N}
 &\multicolumn{1}{c}{T} 
 &\multicolumn{2}{c}{F-test}
 & \multicolumn{1}{c}{$\overline\psi$}\\
 & dd/mm/yyyy & &{(hr)}&{$F_{c}(0.99)$}&{$F_{LC}$} & (\%)&\\
    (1)             &   (2)      &(3)      &(4)  & (5)    &(6)           &(7)\\
 \hline\hline\\

BL J064828.5-174405    & 2018-12-12  &   46  &   2.96  &   2.02  &   3.27   &   9 \\
 \\ 
 BL J071355.7+193500    & 2019-11-08  &   68  &   2.18  &   1.78  &   1.94   &   19 \\
 \\ 
 BL J071355.7+193500    & 2019-12-18  &   14  &   2.53  &   3.91  &   21.61   &   28 \\
 \\ 
BL J071355.7+193500    & 2019-12-19  &   19  &   3.90  &   3.13  &   23.99   &   35 \\
 \\
BL J074836.1+240024    & 2020-01-01  &   18  &   3.88  &   3.24  &   5.54 &   9 \\
 \\ 
BL J081126.7+014652    & 2020-01-04  &   19  &   3.63  &   3.13  &   5.12  &   8 \\
 \\ 
RQQ J082443.8+201901    & 2019-11-08  &   68  &   2.18  &   1.78  &   1.81   &   13 \\
\\
RQQ J085903.6+123903    & 2019-11-26  &   59  &   2.06  &   1.86  &   5.84   &   14 \\
\\
RQQ J092727.4-020832    & 2019-02-12  &   18  &   3.59  &   3.24  &   3.80   &   15 \\
 \\ 
 \\                   
 \hline\hline\\
    \end{tabular}
    \label{tab1:var_result}
    \end{minipage}
    \end{adjustwidth}
\end{table*}

\section{Discussion}
\label{sec:discussion}
The small number of the INOV events confirmed here in the ZTF LCs of the matched samples of blazars and RQQs can probably be ascribed to a combination of (i) the modest photometric sensitivity, specially for the large fraction of objects in our sample that are relatively faint ($m_{r} > $17.5), and (ii) the short durations (T) of the LCs (median $\sim$3.7 hr, Fig.~\ref{fig:sample_dur_distribution})\footnote{It is known that the probability of INOV detection increases with the monitoring duration (T), at least up to 5 - 6 hours (e.g., \citealt{2005A&A...440..855G, 2007AJ....133..303C}).}. Since the duration of nearly all our light curves are under 4 hours, it is likely that the number of INOV detection has been underestimated for this reason. Furthermore, a good fraction of the blazars in the sample, although having a flat radio spectrum, may well be lacking a
high optical polarisation, which is a crucial diagnostic for INOV detection, as shown by \citet{2012A&A...544A..37G}. Nonetheless, the small numbers of INOV events does not detract from the basic objective of our study, which is to look for any exceptionally large, rare INOV events, taking  advantage of the extraordinarily large temporal coverage afforded by the ZTF database. Such an INOV search is particularly salient for RQQs, as compared to blazars for which an extensive INOV literature has accumulated over the past 3 decades. \par
For the blazars, $\psi > 10\%$ stands confirmed in 4 sessions, which is not unfamiliar \citep[see, e.g., the surveys by ][]{1998A&A...329..853H, 1999A&AS..135..477R,2004MNRAS.350..175S,2004MNRAS.348..176S,2021Galax...9..114W}. On the other hand, INOV events of such large amplitudes are exceptional in the case of RQQs, considering their much more modest INOV levels reported in the literature so far (\citealt{1993MNRAS.262..963G,1995MNRAS.274..701G,2003ApJ...586L..25G, 1995ApJ...452..582J, 1997AJ....114..565J, 1999A&AS..135..477R, 2004JApA...25....1S, 2005A&A...440..855G, 2004A&A...421...83R, 2009AJ....138..991R, 2007AJ....133..303C, 2011MNRAS.412.2717J}; \citealt{2012A&A...544A..37G, 2013MNRAS.429.1717J}). These studies have led to an upper limit of $\psi$ = 3 - 5\% for RQQs (GW18). The present work has revealed three INOV sessions with $\psi$ > 10\% (upto 15\%, see Fig.~\ref{fig:light_curve_RQQ_1}). These are J082443.8+201901, J085903.6+123903, and J092727.4-020832 (Table~\ref{tab1:var_result}; Fig.~\ref{fig:light_curve_RQQ_1}). We note that all these quasars are undetected in the VLASS survey at 3 GHz and we have adopted an upper limit of 384 $\mu$Jy for each of them, being 3 rms noise for the survey
(\citealt{2021ApJS..255...30G}). The corresponding upper limits for the radio-loudness parameter $R_{5GHz}$ are 
2.7, 1.2 and 1.9, respectively, i.e. below the threshold R = 10 \citep{1989AJ.....98.1195K}.\footnote{The radio-loudness parameter has been estimated as R =$F_{5GHz}$/$F_{B}$. Due to non-detection of the targets at radio-bands, radio flux ($F_{5GHz}$) has been estimated assuming the flux at 3 GHz to be $< 384 \mu$Jy (i.e., 3$\sigma$ limit of the VLASS survey at 3 GHz, see \citealt{2021ApJS..255...30G}) and scaling it to 5 GHz, using a spectral index of $\alpha = -0.7$. The flux at B-band ($F_{B}$) has been estimated from the apparent magnitude B taken from SIMBAD, using relation given by \citet{1983ApJ...269..352S}. Due to the non availability of $m_{B}$ for the RQQ J092727.4-020832, its B-mag has been estimated from {\it u} and {\it g} magnitudes, using the transformation equations given by \citet{2005AJ....130..873J}.}
Thus we find here that even RQQs can, in rare instances, exhibit INOV amplitudes at levels which had hitherto been recorded only for radio-loud AGN of blazar type. It is also interesting to note that in two of the three sessions (2019-11-08 and 2019-11-16, Fig.~\ref{fig:light_curve_RQQ_1}), the observed INOV is a dip in the LC. These temporally well-resolved emission dips on time scale of $\sim 1$ hour are similar to the striking 4\% amplitude dip of 1-hour duration, which occurred during the intranight optical  monitoring of the TeV blazar PG1553$+$113 \citep{2011MNRAS.416..101G}. Those authors interpreted the short-duration dip in terms of a dusty cloud in the foreground of a superluminal optical knot in the jet, consistent with the scenario of `superluminal gravitational lensing’ proposed for blazar variability \citep{1991Natur.349..766G}. This explanation invoking foreground dusty cloud(s) would also be in accord with the fact that PG 1553+111 is known to have the highest known rotation measure among TeV blazars \citep{2020A&A...634A..87L}. However, if such an explanation were also to be applicable for the emission dips found here in the LCs of the two RQQs (Fig.~\ref{fig:light_curve_RQQ_1}), these RQQs, would also be required to have superluminal optical jets (or, micro-jets) in their nuclei. Thus, even in the event that their observed INOV has an external origin (and not directly originating within a relativistic jets, unlike the INOV of blazars), the present INOV observations of the two RQQs would still require the presence of a relativistic micro-jet in the nucleus. Various arguments and literature supporting such a possibility are summarised, e.g., in \citet{2019MNRAS.485.3009H}.
\section{Conclusions}
\label{sec:conclusion}
The present study underscores the potential of the ongoing Zwicky Transient Facility (ZTF) survey for detecting rare events of strong  intranight optical variability (INOV) of active galactic nuclei, particularly among RQQs whose INOV is generally known to be mild (no more than a few per cent). To search for stronger INOV levels we have made use of ZTF database for a well-defined sample of 53 blazars (156 sessions) and a matched sample containing 132 RQQs (418 sessions). INOV was thus detected in 6 and 3 ZTF sessions, targeting the blazar and RQQ samples, respectively. In these sessions, INOV amplitudes $\psi$ > 10\% were detected not only for the 6 blazar sessions (which is not exceptional) but, unexpectedly, also in three RQQ sessions. This shows that a blazar-like INOV level can also be attained by RQQs, albeit very rarely. Note that two of these 3 INOV events are in fact well-resolved brightness ‘dips’ ( $\sim$ 14\%), lasting about an hour, or less. An analogy with a similar well-resolved dip lasting $\sim$ 1 hour, observed in a light-curve of the TeV blazar PG 1553+111, suggests that the nuclei of these two RQQs probably also contain micro-jets of optical emission performing bulk relativistic motion.
\section*{Acknowledgements}
We thank the referee for various comments and suggestions to improve the quality of the manuscript. We also acknowledge helpful discussions with Prof. Ram Sagar.
GK would like to thank the Indian National Science Academy for a Senior Scientist position. VN thanks the Indian Institute of Astrophysics for hospitality. This work is based on observations obtained with the Samuel Oschin Telescope 48-inch and the 60-inch Telescope at the Palomar Observatory as part of the Zwicky Transient Facility project. ZTF is supported by the National Science Foundation under Grant No. AST-2034437 and a collaboration including Caltech, IPAC, the Weizmann Institute for Science, the Oskar Klein Center at Stockholm University, the University of Maryland, Deutsches Elektronen-Synchrotron and Humboldt University, the TANGO Consortium of Taiwan, the University of Wisconsin at Milwaukee, Trinity College Dublin, Lawrence Livermore National Laboratories, and IN2P3, France. Operations are conducted by COO, IPAC, and UW.

This research has made use of the NASA/IPAC Extragalactic Database (NED) which is operated by the Jet Propulsion Laboratory, California Institute of Technology, under contract with the National Aeronautics and Space Administration.

\section*{Data Availability}
The data used in this study is publicly available in the ZTF DR10.
\bibliographystyle{mnras}
\bibliography{example} 


\newpage
\appendix
\section{Some extra material}
\newpage
\begin{figure*}
    \centering
    \setlength{\fboxsep}{1pt}
    \fbox{\includegraphics[width=0.45\textwidth]{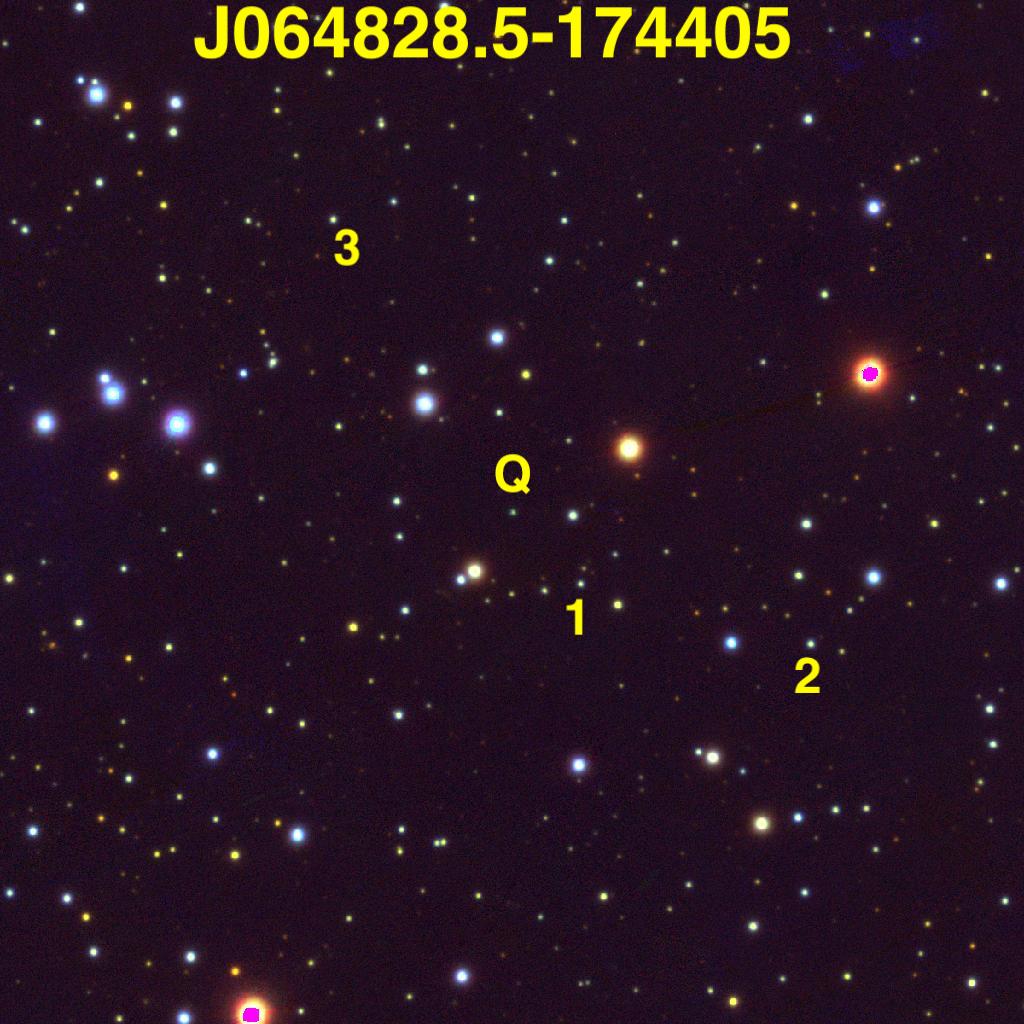}}
    \hfill
    \fbox{\includegraphics[width=0.45\textwidth]{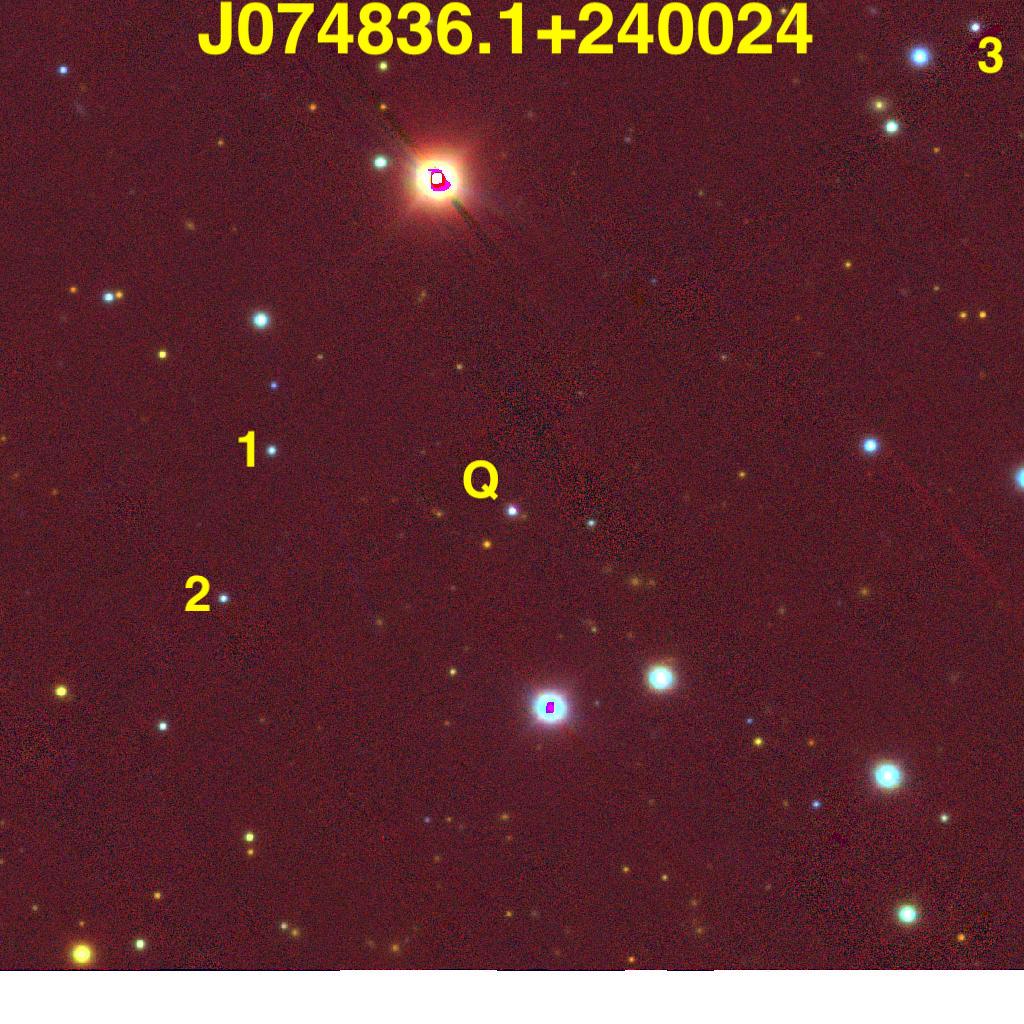}}
    \centering
    \setlength{\fboxsep}{1pt}
    \fbox{\includegraphics[width=0.45\textwidth]{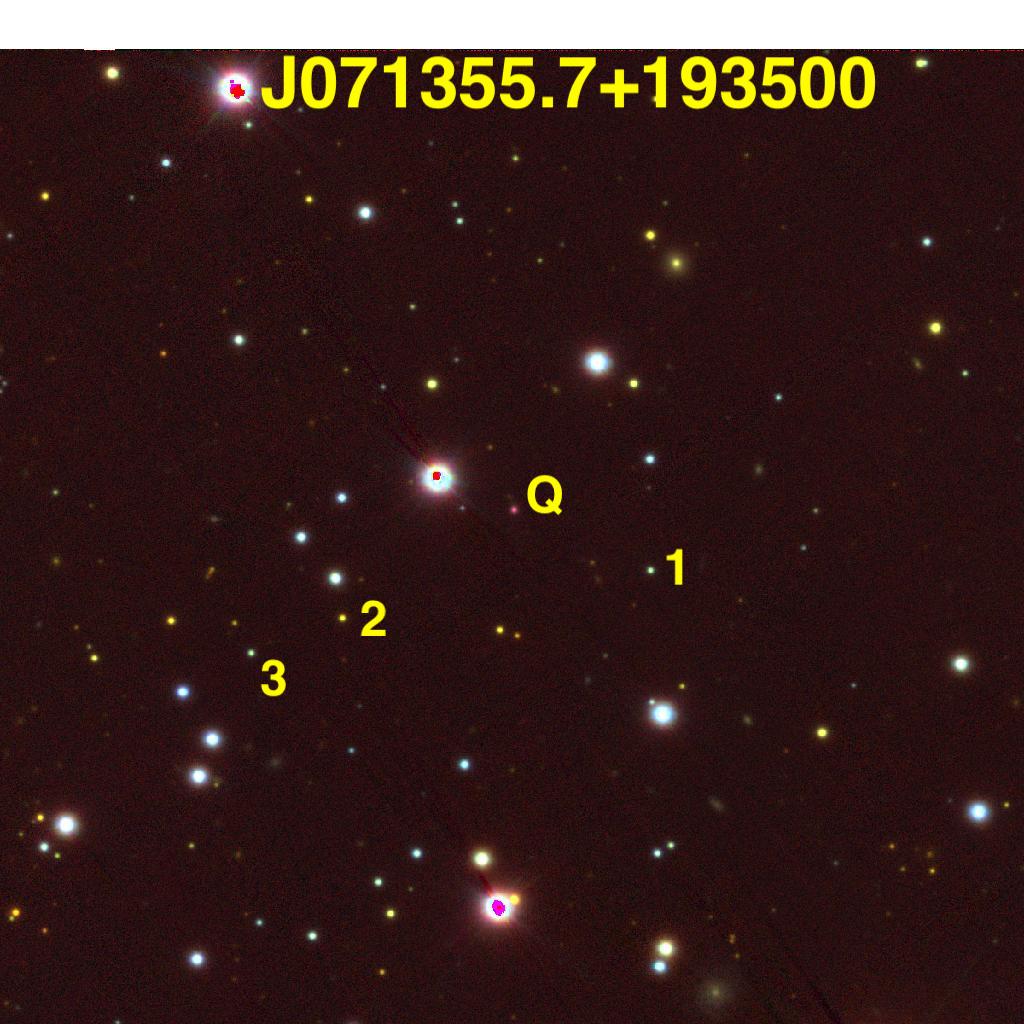}}
    \hfill
    \setlength{\fboxsep}{1pt}
    \fbox{\includegraphics[width=0.45\textwidth]{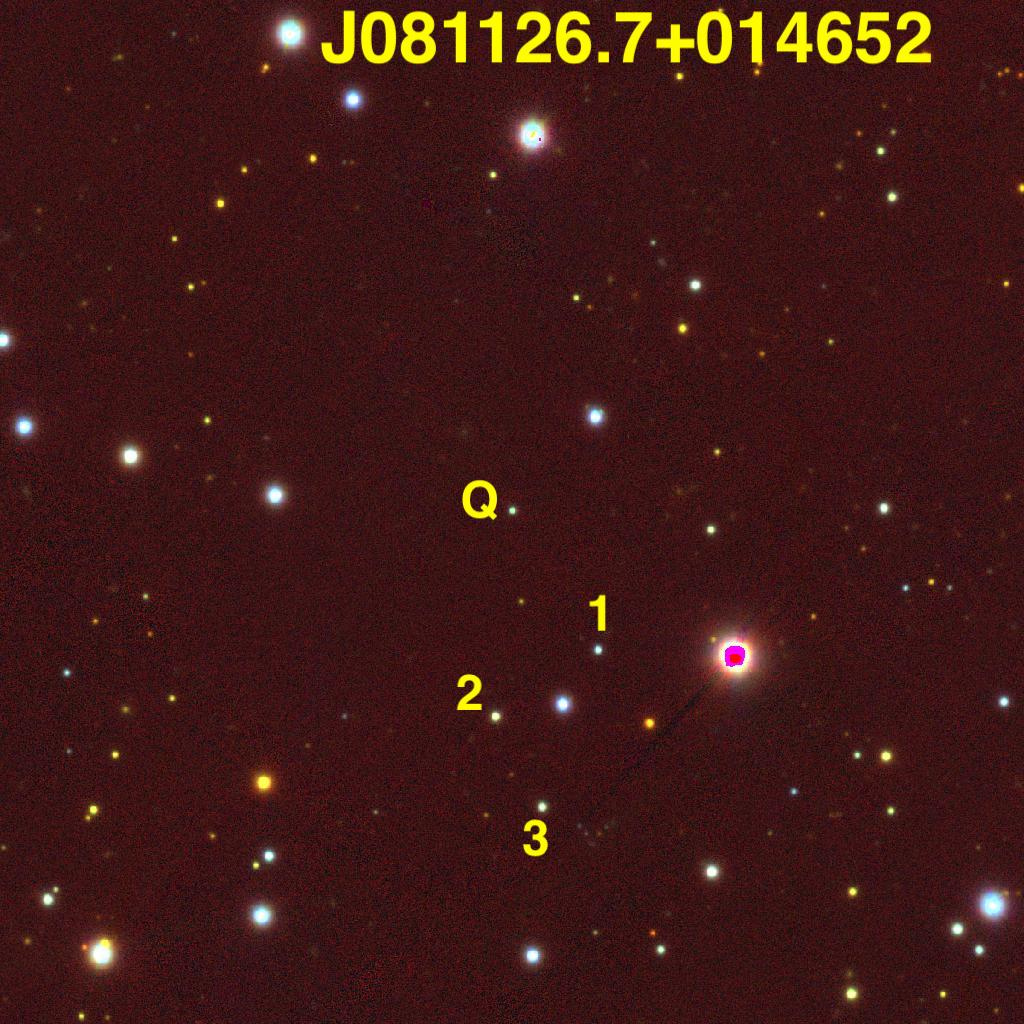}}
    \caption{The $5\arcmin \times 5\arcmin$ optical cutouts of the fields around the four blazars with INOV detection (present work). These cutouts have been taken from the Panstarrs survey \citep{2016arXiv161205560C}. The target blazar and the selected comparison stars are marked with Q, 1, 2, and 3, respectively.}
    \label{fig:final_blazar_image_cutouts}
\end{figure*}
\begin{figure*}
    \centering
    \setlength{\fboxsep}{1pt}
    \fbox{\includegraphics[width=0.45\textwidth]{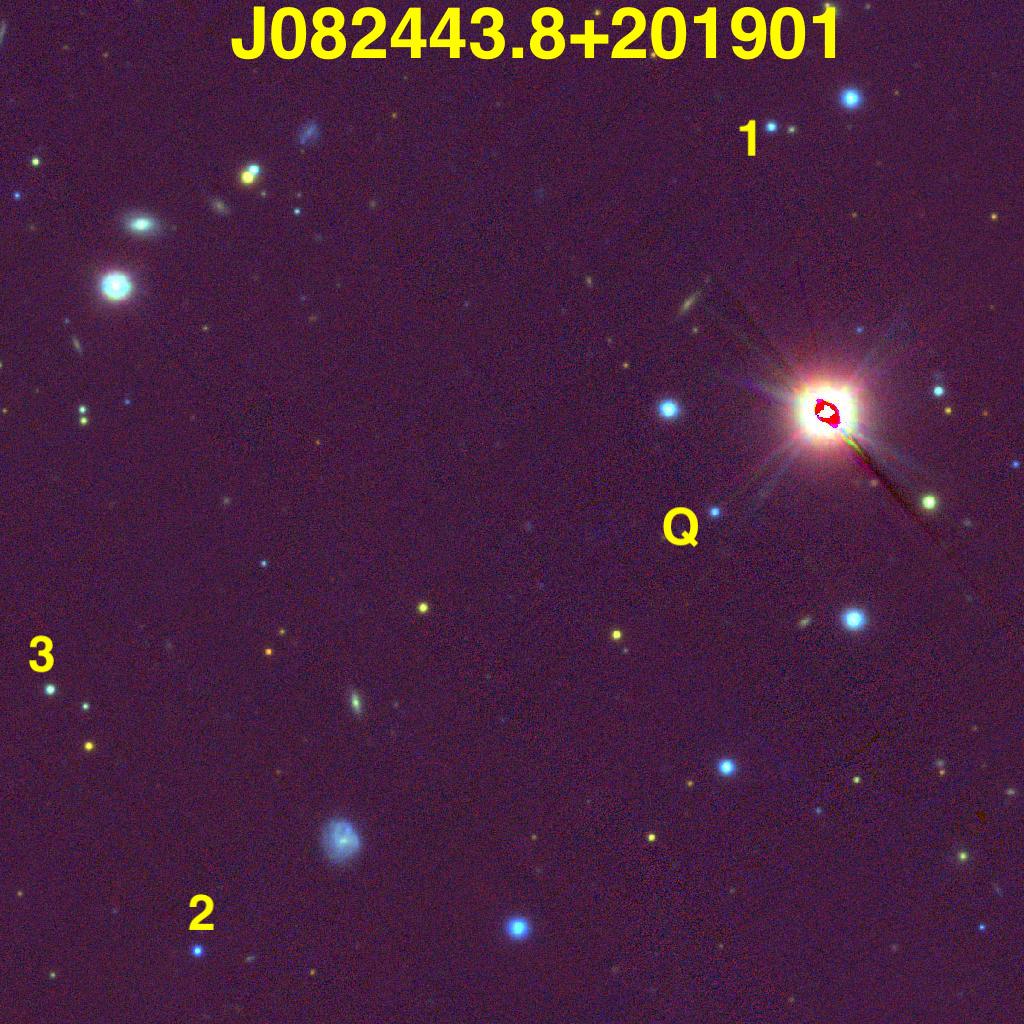}}
    \hfill
    \fbox{\includegraphics[width=0.45\textwidth]{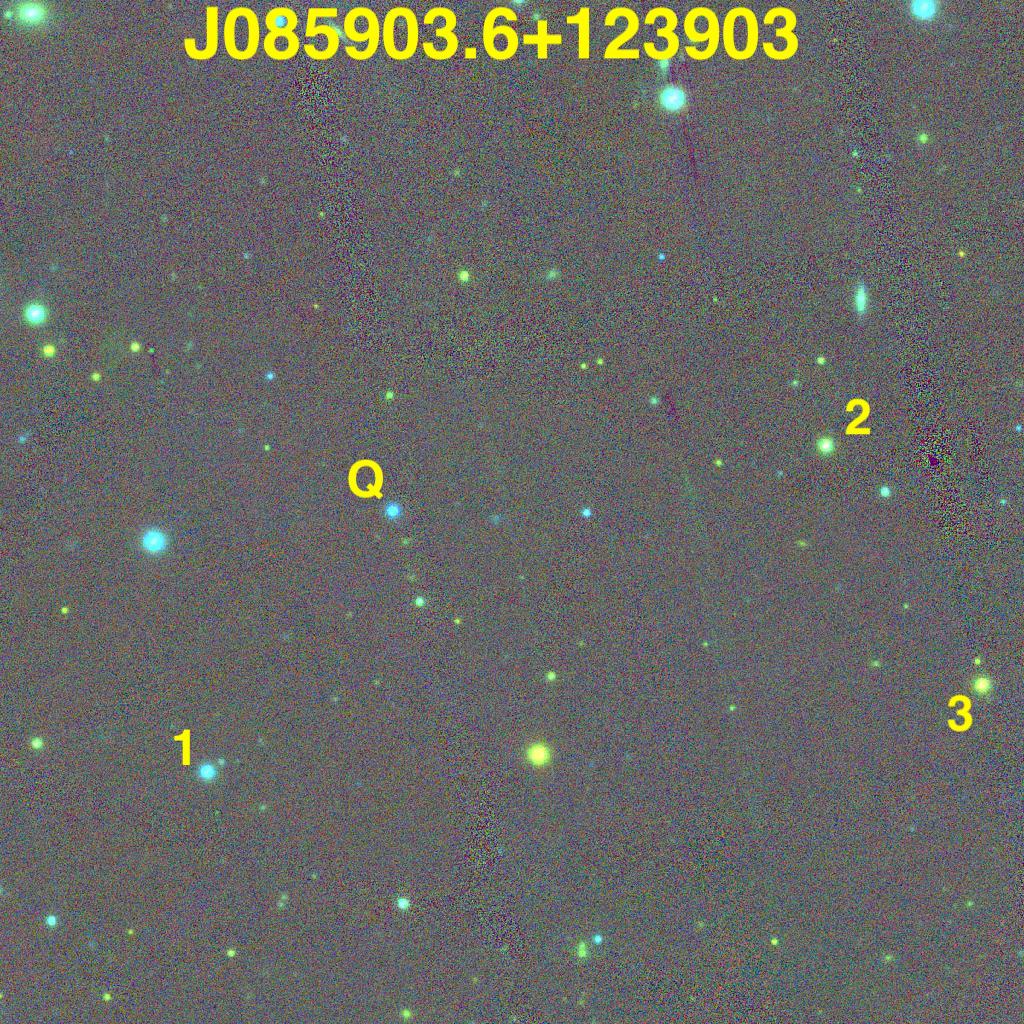}}
    \label{fig:final_rqq_image_cutouts}
\end{figure*}

\begin{figure*}
    \setlength{\fboxsep}{1pt}
    \fbox{\includegraphics[width=0.45\textwidth]{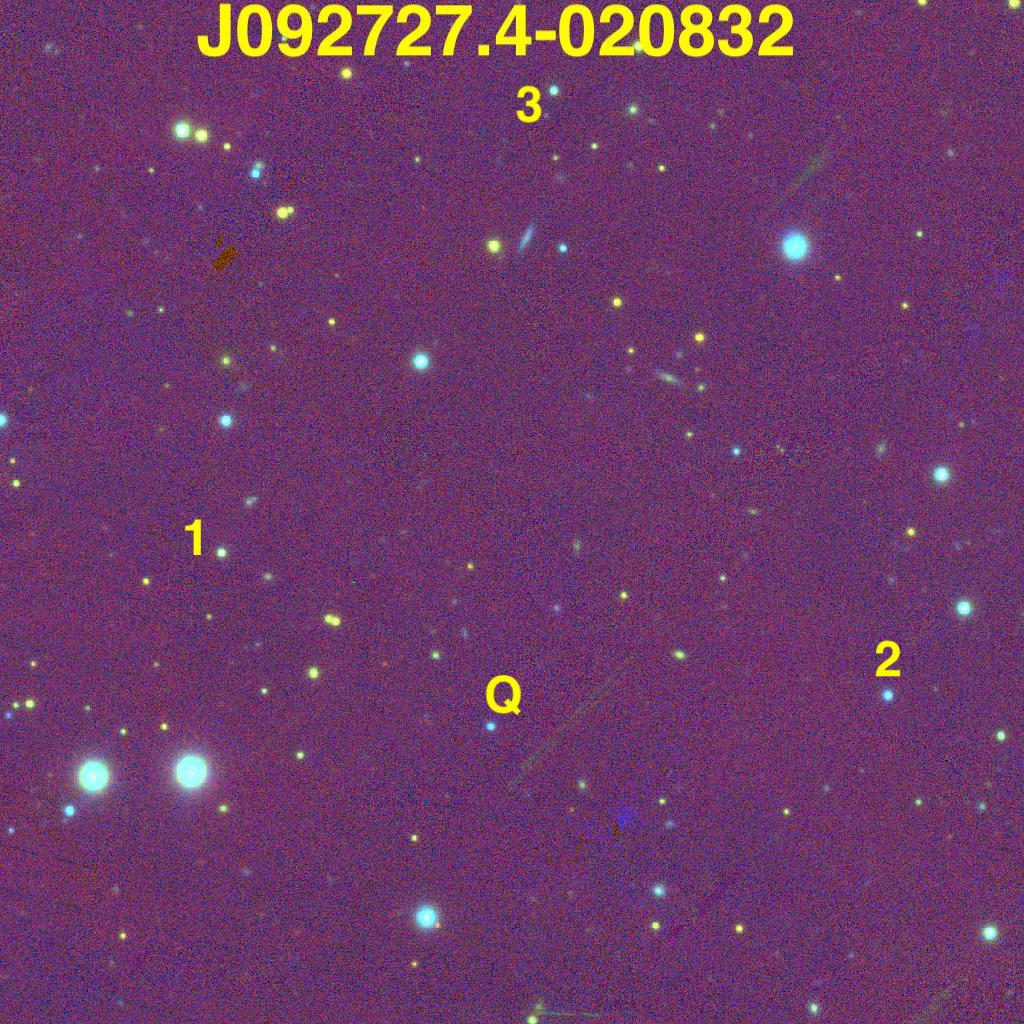}}
    \setcounter{figure}{1}
    \caption{The $5\arcmin \times 5\arcmin$ optical cutouts of the fields around the three RQQs wih INOV detection (present work). These cutouts have been taken from the Panstarrs survey \citep{2016arXiv161205560C}. The target RQQ and the selected comparison stars are marked with Q, 1, 2, and 3, respectively.}
    \label{fig:final_rqq_image_cutouts_2}
\end{figure*}
\vspace{10.0in}
\onecolumn
\begin{longtable}{cccccccc}
\caption{The sample of 53 blazars used in our study.\label{foo}}\\\hline
  \multicolumn{1}{c}{Target Id}  &
  \multicolumn{1}{c}{SDSS name}  & \multicolumn{1}{c}{RA (J2000)} & \multicolumn{1}{c}{Dec (J2000)} & \multicolumn{1}{c}{Redshift} & \multicolumn{1}{c} {$m_{r} ^{\dagger}$}    & \multicolumn{1}{c} {Fermi detection}\\ 
 					 & 		& (deg)    & (deg)    & $(z)$ 	&  & and class$^{\ast}$\\
 \hline
\endfirsthead
\multicolumn{3}{@{}l}{\ldots continued}\\\hline
  \multicolumn{1}{c}{Target Id}  &
  \multicolumn{1}{c}{SDSS name}  & \multicolumn{1}{c}{RA (J2000)} & \multicolumn{1}{c}{Dec (J2000)} & \multicolumn{1}{c}{Redshift} & \multicolumn{1}{c} {$r_{mag}$}    & \multicolumn{1}{c} {Fermi association}\\ 
 					 & 		& (deg)    & (deg)    & $z$ 	&  & \\
\hline
\endhead 
\hline
\multicolumn{3}{r@{}}{continued \ldots}\\
\endfoot
\hline
\endlastfoot
1    &   J054407-224109    &   86.031531  &   -22.686088  &   1.54  &   16.14  &   --- \\
 2    &   J061038-230145    &   92.661609  &   -23.029389  &   2.85  &   18.52  &   --- \\
 3    &   J061357+130645    &   93.490388  &   13.112623  &   0.74  &   18.47  &   --- \\
 4    &   J063053-132334    &   97.724612  &   -13.392908  &   1.02  &   18.28  &   --- \\
 5    &   J064828-174405    &   102.118740  &   -17.734838  &   1.23  &   17.26  &   --- \\
 6    &   J065358+370540    &   103.492837  &   37.094618  &   1.98  &   18.17  &   --- \\
 7    &   J071044+422054    &   107.684692  &   42.348643  &   1.16  &   18.21  &   --- \\
 8    &   J071355+193500    &   108.482003  &   19.583465  &   0.54  &   18.68  &   LSP \\
 9    &   J071424+353439    &   108.603405  &   35.577737  &   1.62  &   18.22  &   --- \\
 10    &   J072201+372228    &   110.505261  &   37.374635  &   1.63  &   17.99  &   --- \\
 11    &   J072516+142513    &   111.320038  &   14.420509  &   1.04  &   18.22  &   LSP \\
 12    &   J072614+215320    &   111.559419  &   21.888938  &   1.86  &   18.46  &   --- \\
 13    &   J073051+404950    &   112.713957  &   40.830805  &   2.50  &   18.50  &   --- \\
 14    &   J073256+254838    &   113.234483  &   25.810795  &   1.44  &   18.60  &   --- \\
 15    &   J073807+174219    &   114.530811  &   17.705291  &   0.42  &   17.21  &   LSP \\
 16    &   J074110+311200    &   115.294601  &   31.200086  &   0.63  &   16.56  &   --- \\
 17    &   J074344+232838    &   115.937384  &   23.477522  &   0.78  &   18.58  &   --- \\
 18    &   J074451+292006    &   116.214022  &   29.335029  &   1.18  &   17.00  &   --- \\
 19    &   J074559+331334    &   116.497199  &   33.226175  &   0.61  &   17.91  &   --- \\
 20    &   J074836+240024    &   117.150463  &   24.006714  &   0.41  &   17.87  &   LSP \\
 21    &   J075052+123104    &   117.716864  &   12.518025  &   0.89  &   17.04  &   --- \\
 22    &   J075448+303355    &   118.703559  &   30.565299  &   0.80  &   18.02  &   --- \\
 23    &   J075650-154205    &   119.211253  &   -15.701501  &   1.42  &   18.75  &   --- \\
 24    &   J080018+164556    &   120.078301  &   16.765915  &   0.31  &   18.17  &   --- \\
 25    &   J081126+014652    &   122.861289  &   1.781181  &   1.15  &   18.02  &   LSP \\
 26    &   J082550+030924    &   126.459751  &   3.156819  &   0.51  &   18.74  &   LSP \\
 27    &   J084205+183541    &   130.521237  &   18.594738  &   1.27  &   16.79  &   --- \\
 28    &   J085317+281350    &   133.324298  &   28.230585  &   0.92  &   18.04  &   --- \\
 29    &   J085441+575729    &   133.674992  &   57.958341  &   1.32  &   17.49  &   --- \\
 30    &   J085448+200630    &   133.703651  &   20.108531  &   0.31  &   15.09  &   LSP \\
 31    &   J090910+012135    &   137.292055  &   1.359905  &   1.02  &   17.07  &   LSP \\
 32    &   J090924+521632    &   137.352847  &   52.275750  &   0.41  &   17.68  &   --- \\
 33    &   J091147+334916    &   137.949012  &   33.821361  &   0.46  &   18.71  &   HSP \\
 34    &   J092058+444153    &   140.243586  &   44.698345  &   2.19  &   17.83  &   LSP \\
 35    &   J092223-052907    &   140.598640  &   -5.485319  &   0.97  &   17.45  &   LSP \\
 36    &   J092507+001913    &   141.282581  &   0.320545  &   1.72  &   17.15  &   --- \\
 37    &   J092824+444604    &   142.100609  &   44.768022  &   1.90  &   18.32  &   --- \\
 38    &   J093035+464408    &   142.646174  &   46.735759  &   2.03  &   18.53  &   --- \\
 39    &   J093309+461535    &   143.288817  &   46.259857  &   0.78  &   18.42  &   --- \\
 40    &   J095227+504850    &   148.113797  &   50.814097  &   1.09  &   17.68  &   --- \\
 41    &   J095738+552257    &   149.409102  &   55.382734  &   0.90  &   17.28  &   LSP \\
 42    &   J100724+580203    &   151.853701  &   58.034324  &   3.35  &   17.49  &   --- \\
 43    &   J100811+470521    &   152.047613  &   47.089325  &   0.34  &   18.42  &   --- \\
 44    &   J104146+523328    &   160.444922  &   52.557857  &   0.68  &   17.04  &   --- \\
 45    &   J104703-130832    &   161.766388  &   -13.142327  &   1.29  &   17.21  &   --- \\
 46    &   J183243+135747    &   278.179821  &   13.963350  &   2.83  &   17.23  &   --- \\
 47    &   J185027+282513    &   282.614955  &   28.420337  &   2.54  &   17.49  &   --- \\
 48    &   J191810+552038    &   289.544790  &   55.344070  &   1.73  &   18.84  &   --- \\
 49    &   J193653-040245    &   294.223937  &   -4.045945  &   0.49  &   18.66  &   --- \\
 50    &   J195542+513148    &   298.928072  &   51.530154  &   1.21  &   18.08  &   LSP \\
 51    &   J202456+171813    &   306.235685  &   17.303681  &   1.05  &   17.83  &   --- \\
 52    &   J203147+545503    &   307.949834  &   54.917550  &   1.26  &   17.07  &   --- \\
 53    &   J235342+551840    &   358.426244  &   55.311306  &   1.93  &   18.47  &   --- \\
   \hline
  \multicolumn{7}{{p{0.80\columnwidth}}}{\textbf{$^{\ast}$} LSP: Low Synchrotron Peaked blazar (i.e., $\nu_{s}^{peak} < 10^{14}$ Hz), ISP: Intermediate Synchrotron Peaked blazar (i.e., $10^{14} \leq \nu_{s}^{peak} \leq 10^{15}$ Hz), HSP: High Synchrotron Peaked blazar (i.e., $\nu_{s}^{peak} > 10^{16}$ Hz) \citep{2010ApJ...716...30A};  {\bf $\dagger$} Median magnitude over the ZTF / INOV session.}
\label{tab:sample_table_BL}
\end{longtable}

\newpage

\begin{longtable}{cccccccc}
\caption{The sample of 132 RQQs used in our study.\label{foo}}\\\hline
  \multicolumn{1}{c}{Target Id}  &
  \multicolumn{1}{c}{SDSS name}  & \multicolumn{1}{c}{RA (J2000)} & \multicolumn{1}{c}{Dec (J2000)} & \multicolumn{1}{c}{Redshift} & \multicolumn{1}{c} {$m_{r}^{\dagger}$}\\ 
 					 & 		& (deg)    & (deg)    & $z$ 	&  & \\
 \hline
\endfirsthead
\multicolumn{3}{@{}l}{\ldots continued}\\\hline
  \multicolumn{1}{c}{Target Id}  &
  \multicolumn{1}{c}{SDSS name}  & \multicolumn{1}{c}{RA (J2000)} & \multicolumn{1}{c}{Dec (J2000)} & \multicolumn{1}{c}{Redshift} & \multicolumn{1}{c} {$m_{r}^{\dagger}$}\\ 
 					 & 		& (deg)    & (deg)    & $(z)$ 	& \\
\hline
\endhead 
\hline
\multicolumn{3}{r@{}}{continued \ldots}\\
\endfoot
\hline
\endlastfoot
1    &   J032108+413220    &   50.285212  &   41.539155  &   2.47  &   17.12 \\
 2    &   J072517+434553    &   111.323039  &   43.764857  &   1.59  &   17.53 \\
 3    &   J073330+370818    &   113.378585  &   37.138522  &   1.14  &   18.09 \\
 4    &   J074125+435605    &   115.354670  &   43.934930  &   0.66  &   17.91 \\
 5    &   J074224+342847    &   115.604057  &   34.480018  &   1.00  &   18.28 \\
 6    &   J074358+323512    &   115.992811  &   32.586832  &   0.91  &   17.25 \\
 7    &   J074420+381839    &   116.086562  &   38.311087  &   0.41  &   17.81 \\
 8    &   J074445+200014    &   116.190399  &   20.004137  &   1.72  &   17.03 \\
 9    &   J074451+292006    &   116.214022  &   29.335029  &   1.18  &   16.92 \\
 10    &   J074458+301849    &   116.243619  &   30.313655  &   0.58  &   17.90 \\
 11    &   J074511+191942    &   116.296970  &   19.328622  &   0.38  &   18.17 \\
 12    &   J074527+222453    &   116.364277  &   22.414816  &   1.35  &   17.30 \\
 13    &   J074601+255635    &   116.507260  &   25.943216  &   0.68  &   17.91 \\
 14    &   J074645+314149    &   116.687709  &   31.697041  &   0.33  &   17.86 \\
 15    &   J074820+340752    &   117.087376  &   34.131323  &   0.34  &   18.15 \\
 16    &   J075054+425219    &   117.727694  &   42.872028  &   1.91  &   16.19 \\
 17    &   J075222+273823    &   118.095486  &   27.639782  &   1.06  &   17.06 \\
 18    &   J075251+181108    &   118.212981  &   18.185625  &   1.92  &   18.17 \\
 19    &   J075331+182117    &   118.382698  &   18.354798  &   1.83  &   18.71 \\
 20    &   J075524+342134    &   118.850456  &   34.359598  &   2.12  &   17.80 \\
 21    &   J075545+142941    &   118.941219  &   14.494748  &   0.54  &   18.04 \\
 22    &   J075851+220457    &   119.714674  &   22.082731  &   0.82  &   16.98 \\
 23    &   J080112+191544    &   120.302700  &   19.262468  &   0.41  &   16.73 \\
 24    &   J080141+124208    &   120.423193  &   12.702425  &   1.36  &   18.75 \\
 25    &   J080148+174316    &   120.453578  &   17.721203  &   0.42  &   18.67 \\
 26    &   J080320+252602    &   120.833697  &   25.434016  &   0.48  &   17.68 \\
 27    &   J080356+365723    &   120.986896  &   36.956654  &   1.23  &   18.02 \\
 28    &   J080630+144242    &   121.626243  &   14.711801  &   1.22  &   16.94 \\
 29    &   J080704+360353    &   121.770407  &   36.064887  &   0.47  &   17.69 \\
 30    &   J080954+074355    &   122.476615  &   7.732000  &   0.65  &   16.61 \\
 31    &   J081014+204021    &   122.560971  &   20.672652  &   2.52  &   17.33 \\
 32    &   J081331+254503    &   123.380356  &   25.750871  &   1.51  &   16.08 \\
 33    &   J081558+154055    &   123.993161  &   15.682035  &   2.23  &   17.72 \\
 34    &   J082234+170935    &   125.643085  &   17.160016  &   1.78  &   18.84 \\
 35    &   J082443+201901    &   126.182536  &   20.317028  &   0.54  &   18.75 \\
 36    &   J083103+214553    &   127.765630  &   21.764860  &   1.53  &   18.60 \\
 37    &   J084912+144754    &   132.302033  &   14.798563  &   0.86  &   18.04 \\
 38    &   J084952+363927    &   132.469769  &   36.657773  &   1.22  &   17.51 \\
 39    &   J085435+311020    &   133.649394  &   31.172445  &   0.85  &   17.27 \\
 40    &   J085502+321730    &   133.761865  &   32.291824  &   1.11  &   17.53 \\
 41    &   J085507+110139    &   133.779817  &   11.027697  &   1.07  &   18.28 \\
 42    &   J085903+123903    &   134.764826  &   12.651006  &   1.21  &   17.60 \\
 43    &   J085648+141321    &   134.202769  &   14.222786  &   0.34  &   17.69 \\
 44    &   J085733+160017    &   134.388345  &   16.004941  &   0.83  &   17.27 \\
 45    &   J085813+203746    &   134.555582  &   20.629494  &   0.79  &   17.48 \\
 46    &   J085831+554922    &   134.630476  &   55.822876  &   2.06  &   17.70 \\
 47    &   J085853+063909    &   134.722608  &   6.652667  &   1.89  &   18.47 \\
 48    &   J085924+463717    &   134.851446  &   46.621488  &   0.92  &   16.99 \\
 49    &   J090014+232109    &   135.058991  &   23.352713  &   1.42  &   18.60 \\
 50    &   J090026+204158    &   135.110491  &   20.699672  &   0.71  &   17.91 \\
 51    &   J090219+233534    &   135.581513  &   23.593048  &   1.30  &   18.07 \\
 52    &   J090450+230118    &   136.209173  &   23.021895  &   0.48  &   17.88 \\
 53    &   J090508+074151    &   136.286912  &   7.697576  &   2.04  &   17.72 \\
 54    &   J090529+253033    &   136.371561  &   25.509275  &   0.82  &   18.47 \\
 55    &   J090607+030036    &   136.531985  &   3.010051  &   1.68  &   18.22 \\
 56    &   J090824+033929    &   137.100603  &   3.658321  &   1.26  &   16.89 \\
 57    &   J090832+102152    &   137.135897  &   10.364733  &   0.79  &   17.19 \\
 58    &   J090850+254447    &   137.210097  &   25.746396  &   0.45  &   17.96 \\
 59    &   J090903+473716    &   137.263959  &   47.621156  &   1.04  &   17.48 \\
 60    &   J090906+323630    &   137.275758  &   32.608415  &   0.81  &   17.02 \\
 61    &   J090916+163522    &   137.317023  &   16.589605  &   1.07  &   17.47 \\
 62    &   J091120+172231    &   137.836808  &   17.375582  &   0.65  &   17.91 \\
 63    &   J091135+293803    &   137.895866  &   29.634379  &   0.61  &   17.90 \\
 64    &   J091145+145556    &   137.937671  &   14.932477  &   1.27  &   17.17 \\
 65    &   J091213+513210    &   138.057214  &   51.536264  &   1.48  &   18.75 \\
 66    &   J091307+442014    &   138.282660  &   44.337322  &   2.93  &   18.51 \\
 67    &   J091334+352540    &   138.395716  &   35.428052  &   2.16  &   17.85 \\
 68    &   J091431+083742    &   138.632417  &   8.628568  &   0.65  &   17.08 \\
 69    &   J091511+201248    &   138.795985  &   20.213429  &   1.24  &   17.15 \\
 70    &   J091624+040943    &   139.103242  &   4.162055  &   0.31  &   17.81 \\
 71    &   J091755+053749    &   139.479292  &   5.630493  &   0.35  &   18.19 \\
 72    &   J091811+524255    &   139.545982  &   52.715538  &   2.11  &   17.89 \\
 73    &   J091845+060226    &   139.691350  &   6.040617  &   0.79  &   17.51 \\
 74    &   J091938+360246    &   139.911014  &   36.046415  &   0.60  &   17.90 \\
 75    &   J092011+495403    &   140.047255  &   49.901111  &   1.69  &   17.12 \\
 76    &   J092100+183605    &   140.251579  &   18.601638  &   1.24  &   17.69 \\
 77    &   J092234+204020    &   140.644077  &   20.672439  &   0.94  &   18.04 \\
 78    &   J092311+450918    &   140.796077  &   45.155261  &   1.11  &   18.03 \\
 79    &   J092537+102115    &   141.406856  &   10.354286  &   1.15  &   17.55 \\
 80    &   J092635+044013    &   141.646336  &   4.670418  &   1.43  &   17.61 \\
 81    &   J092658+093248    &   141.745314  &   9.546807  &   0.42  &   17.97 \\
 82    &   J092702+084758    &   141.760914  &   8.799665  &   2.20  &   17.88 \\
 83    &   J092703+522316    &   141.763298  &   52.387948  &   0.60  &   16.74 \\
 84    &   J092712+124458    &   141.801348  &   12.749728  &   1.28  &   17.00 \\
 85    &   J092727-020833    &   141.864283  &   -2.142488  &   0.77  &   18.57 \\
 86    &   J092756+253007    &   141.987012  &   25.502194  &   1.20  &   16.83 \\
 87    &   J093015+482830    &   142.564691  &   48.475261  &   0.50  &   18.76 \\
 88    &   J093021+235329    &   142.591111  &   23.891627  &   1.17  &   17.12 \\
 89    &   J093046+031326    &   142.693573  &   3.224156  &   1.48  &   17.62 \\
 90    &   J093048+230637    &   142.703793  &   23.110538  &   0.47  &   18.68 \\
 91    &   J093113+052421    &   142.806463  &   5.406035  &   1.25  &   17.34 \\
 92    &   J093131+100337    &   142.879506  &   10.060509  &   0.80  &   18.47 \\
 93    &   J093210+501028    &   143.045254  &   50.174574  &   0.51  &   17.96 \\
 94    &   J093244+500938    &   143.184913  &   50.160668  &   0.84  &   18.47 \\
 95    &   J093518+020415    &   143.825816  &   2.071010  &   0.65  &   16.99 \\
 96    &   J093520+335548    &   143.836138  &   33.930054  &   0.69  &   17.91 \\
 97    &   J093531+451927    &   143.882227  &   45.324364  &   0.67  &   18.98 \\
 98    &   J093623+171726    &   144.099697  &   17.290848  &   0.50  &   18.76 \\
 99    &   J093651+104611    &   144.213051  &   10.769878  &   2.77  &   18.55 \\
 100    &   J093759+453801    &   144.498376  &   45.633888  &   0.43  &   18.75 \\
 101    &   J093809+450420    &   144.539700  &   45.072382  &   1.55  &   17.51 \\
 102    &   J093826+192935    &   144.608604  &   19.493153  &   0.40  &   18.03 \\
 103    &   J093833+225031    &   144.641444  &   22.842134  &   0.43  &   17.69 \\
 104    &   J094202+042244    &   145.508538  &   4.379059  &   3.27  &   17.40 \\
 105    &   J094323+020411    &   145.847240  &   2.069836  &   0.95  &   18.14 \\
 106    &   J094404+480646    &   146.018439  &   48.112955  &   0.39  &   18.17 \\
 107    &   J094526+083629    &   146.361811  &   8.608149  &   1.46  &   18.75 \\
 108    &   J094607+495412    &   146.530444  &   49.903581  &   0.76  &   17.48 \\
 109    &   J094617+135425    &   146.574110  &   13.907049  &   1.85  &   18.71 \\
 110    &   J094651+494703    &   146.713263  &   49.784333  &   0.99  &   18.05 \\
 111    &   J094754+223752    &   146.975601  &   22.631240  &   2.54  &   18.58 \\
 112    &   J095047+480047    &   147.697857  &   48.013174  &   1.74  &   17.17 \\
 113    &   J095122+223144    &   147.843926  &   22.528937  &   2.22  &   17.86 \\
 114    &   J095216+204316    &   148.068856  &   20.721302  &   1.43  &   17.25 \\
 115    &   J095411+561655    &   148.549958  &   56.282195  &   2.10  &   17.79 \\
 116    &   J095714+544017    &   149.311138  &   54.671553  &   2.59  &   17.43 \\
 117    &   J095852+465028    &   149.717062  &   46.841349  &   1.16  &   18.02 \\
 118    &   J095856+570820    &   149.734265  &   57.138970  &   1.45  &   18.75 \\
 119    &   J100114+470232    &   150.309700  &   47.042271  &   2.17  &   17.72 \\
 120    &   J100133+570839    &   150.391110  &   57.144226  &   1.37  &   17.29 \\
 121    &   J100336+460550    &   150.902763  &   46.097445  &   1.66  &   17.99 \\
 122    &   J100502+465927    &   151.259788  &   46.990950  &   0.95  &   17.09 \\
 123    &   J100755+452824    &   151.982807  &   45.473522  &   1.65  &   18.08 \\
 124    &   J101330+531559    &   153.375730  &   53.266567  &   1.51  &   16.31 \\
 125    &   J101719+441704    &   154.330649  &   44.284505  &   0.39  &   17.81 \\
 126    &   J102349+522151    &   155.955783  &   52.364246  &   0.96  &   17.00 \\
 127    &   J103430+470820    &   158.627206  &   47.138934  &   0.78  &   18.03 \\
 128    &   J104157+473329    &   160.491437  &   47.558253  &   1.64  &   17.26 \\
 129    &   J104336+494707    &   160.903063  &   49.785466  &   2.19  &   17.78 \\
 130    &   J104621+483322    &   161.589891  &   48.556317  &   1.58  &   18.22 \\
 131    &   J105043+500329    &   162.683299  &   50.058170  &   0.62  &   17.51 \\
 132    &   J105141+523934    &   162.922480  &   52.659525  &   0.98  &   18.28 \\\\
 \hline
  \multicolumn{7}{{p{0.80\columnwidth}}}{{\bf $\dagger$:} Median magnitude over the ZTF / INOV session.}
\label{tab:sample_table_RQQ}
\end{longtable}
\bsp	
\label{lastpage}
\end{document}